\documentclass[aps,prd,superscriptaddress,showpacs,twocolumn,preprintnumbers,showkeys,superscriptaddress,amsmath,amssymb,nofootinbib]{revtex4-2}
\usepackage{array}
\usepackage{booktabs}
\usepackage{commath}
\usepackage{graphicx}
\usepackage{epstopdf}
\usepackage{amsmath}
\usepackage{amsfonts}
\usepackage{amssymb}
\usepackage{latexsym}
\usepackage{hyperref}
\usepackage[english]{babel}
\usepackage[utf8]{inputenc}
\usepackage[colorinlistoftodos]{todonotes}
\usepackage{xcolor}
\usepackage{slashed}
\usepackage{bm}  
\usepackage[normalem]{ulem} 
\usepackage{amsmath,mathtools}
\usepackage{caption}
\usepackage{subcaption}
\usepackage{bbm}
\usepackage{dsfont}
\usepackage[toc]{appendix}
\usepackage{float}
\useunder{\uline}{\ul}{}
\usepackage{hyperref}
\begin{document}
\title{A Lorentz-violating low-energy model for the bilayer Graphene}
\author{Y. M. P. Gomes}
\email{yurimullergomes@gmail.com}
\affiliation{Departamento de F\'{\i}sica Te\'{o}rica, Universidade do
  Estado do Rio de
Janeiro, 20550-013 Rio de Janeiro, Brazil}
\author{M. J. Neves}\email{mariojr@ufrrj.br}
\affiliation{Departamento de F\'isica, Universidade Federal Rural do Rio de Janeiro,
BR 465-07, 23890-971, Serop\'edica, Rio de Janeiro, Brazil}

\begin{abstract}

In this work, we propose a model with Lorentz symmetry violation which describes the electronic low energy limit of the AA-bilayer graphene (BLG) system. The AA-type bilayer is known to preserve the linear dispersion relation of the graphene layer in the low energy limit. The theoretical model shows that in the BLG system, a time-like vector can be associated with the layer separation and contributes to the energy eigenstates. Based on these properties, we can describe in a $(2+1)$-dimensional space-time the fermionic quasi-particles that emerge in the low-energy limit with the introduction of a Lorentz-violating parameter, in analogy with the $(3 + 1)$-dimensional Standard Model Extension (SME). Moreover, we study the consequences of the coupling of these fermionic quasi-particles with the electromagnetic field, and we show via effective action that the low-energy photon acquires a massive spectrum. Finally, using the hydrodynamic approach in the collisionless limit, one finds that the LSV generates a new kind of anomalous thermal current to the vortexes of the system via coupling of the LSV vector.

\end{abstract}

\maketitle

\pagestyle{myheadings}


\section{Introduction}
In the last decades, the theoretical study of high energy physics reveals that in some unification theories can arise a possible violation of the Lorentz symmetry \cite{intro1, intro2}. Based on these results, the Lorentz symmetry violation (LSV) has been intensively sought over the last decade, for instance in the energy spectrum of hydrogen, in the generation of a momentum-dependent electric dipole moment for charged leptons, and in the neutrino's oscillations \cite{intro3, intro4}.
Inspired by these LSV models, more recently, LSV-like models were successfully applied to the condensed matter physics to describe three-dimensional Weyl semi-metals (3DWSM) \cite{intro5,intro6}, in which the introduction of a constant axial four-vector minimally coupled with the electrons explains the low energy spectrum of the 3DWSM. This model has proven to be able to predict the existence of a Carrol-Field-Jackiw (CFJ) term in the electromagnetic response, properly describing the anomalous Hall current, and also it predicts the existence of a chiral anomaly in the 3DWSM system. Non-linear optical properties are studied in semiconductors systems \cite{Chang23,Chang22}. The optical refractive index changes in quantum wells through polaron effects, and are associated with electrons coupled to the phonon \cite{Zhang}. 
This analogy between LSV in high energy physics and condensed matter also influences our understanding of the planar realm. The recent discovery of two-dimensional Weyl semi-metals \cite{hirata} and their theoretical model has shown that the low energy behavior is described by Weyl-like Hamiltonian systems and has quasi-particles that behave like Weyl fermions. These low dimensional systems also have characteristics of anisotropy and tilting of the Dirac cone, and it can be modeled by a $(1+2)$ dimensional LSV lagrangian \cite{Yuri}.
In this work, we propose the description of a graphene bilayer organized at the AA configuration through a fermionic model described by four-component spinors and in the presence of a background three-vector. The AA-stacked bilayer is known to have linear dispersion relation similar to monolayer graphene, in opposition to the AB-stacked graphene which has quadratic dispersion relations. The presence of a sheet near the other one breaks the Lorentz symmetry and its breaking manifests via a constant energy potential which can be modeled through a constant background vector. This vector that explicitly breaks the Lorentz symmetry provides a way to reproduce the low-energy spectrum of the material. Therefore, we study the model from the point of view of a quantum field theory in the presence of the LSV. The fermion model is so minimally coupled to the electromagnetic (EM) field by the abelian gauge symmetry principle in which the perturbative formalism is introduced for a small coupling constant. Thereby, the model contains the pseudo-electrodynamics in $(1+2)$-dimensions \cite{Marino93,Ozela19,Ozela22}, with a time-like LSV parameter in the fermionic sector \cite{Gusynin}. Non-Abelian formulations also are studied in graphene bilayer \cite{sanjose}. The abelian perturbative approach is used to investigate the effects of magnetic fields on the graphene energy spectrum \cite{Kandemir,Dalmazi2000}.  Assuming the smallness of the electromagnetic coupling constant, we calculate
the contributions for the mass of the fermionic quasi-particles, and also for the vacuum polarization tensor at the one-loop approximation.
The consequences of the vacuum polarization in the dynamics of the EM field are investigated in which we discuss the low energy limit.
Posteriorly, we apply this fermion framework in hydrodynamics through the kinetic equation in the presence of a uniform EM background and show the appearance of a new kind of anomalous thermal current via the combination of the LSV parameter with the vortexes of the system. The advantage of our approach compared with numerical calculations is that one can obtain analytical results that could confront or complement the numerical data and also can bring new insights about the system that the numerical assumptions might hide.
The paper is organized as follows: In section \ref{sec2}, the $2D$ Dirac semi-metals are reviewed. The section \ref{sec3}
is due to the bilayer graphene with an LSV fermionic approach for the AA configuration. In section \ref{sec4},
we study the AA configuration coupled to the EM field, and we obtain the vacuum polarization at one loop. The section \ref{sec5}
is dedicated to the contribution of the mass of the fermionic quasi-particles. In section \ref{sec6}, we apply
the LSV fermionic approach in hydrodynamics. For the end, we highlight the conclusions in the section \ref{sec7}.
The results for the loop integrals are shown in the appendix \ref{appendix}.

In this paper, we adopt the natural units in which $\hbar=c=1$, and the Minkowski metric is $\eta^{\mu\nu}=\mbox{diag}(+1,-1,-1)$ in the
$(1+2)$ space-time. One also uses the following nomenclature: $\sigma^\mu$ for the $2 \times 2$ Pauli matrices, Greek letters as $\gamma^\mu$ for the $4 \times 4$ Dirac matrices, and capital Greek letters $\Gamma^\mu$ for the $8 \times 8$ version of the Dirac matrices.

\section{Low-energy Hamiltonian}
\label{sec2}
\begin{figure}[h]
\centering
\includegraphics[scale=0.45]{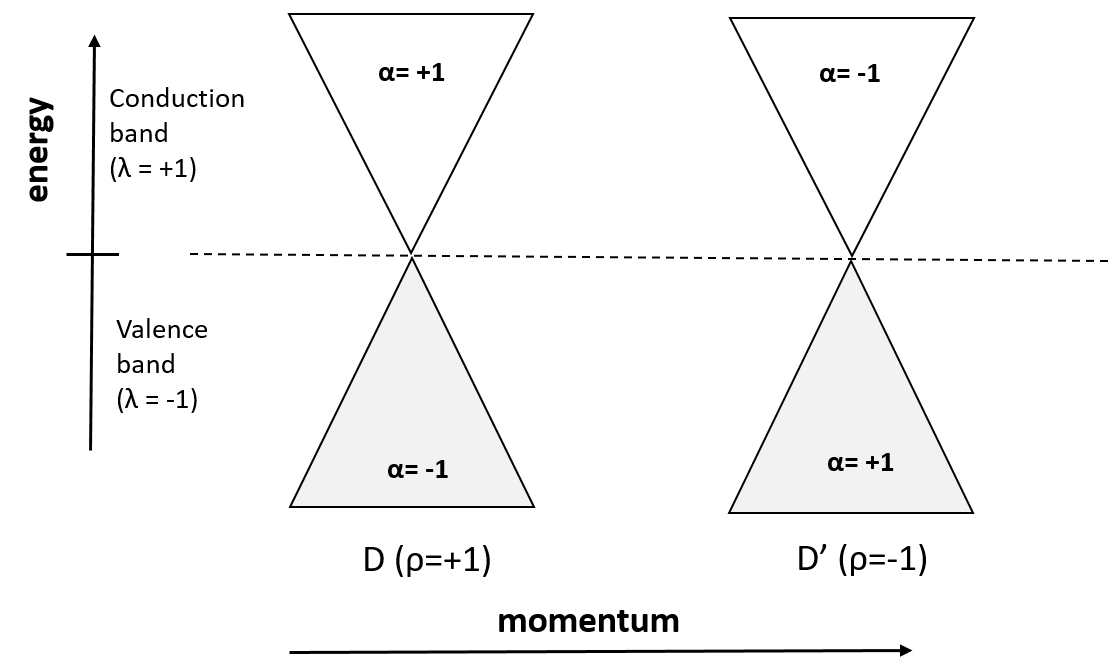}
\caption{The relation between the band index $\lambda$, the valley pseudo-spin $\rho$, and the
chirality $\alpha$ in bidimensional Dirac-like materials.}
\label{fig1}
\end{figure}

The low-energy electron in a single layer of graphene is governed by the Hamiltonian :
\begin{equation}\label{H1}
H(q_x,q_y) =  v_F \left( \, q_x \, \sigma^x + q_y \, \sigma^y \, \right) \; ,
\end{equation}
where $v_F\approx 1/300$ is the Fermi velocity, and $\sigma^x$, $\sigma^y$ are the usual Pauli matrices. By convenience, the $2 \times 2$ identity matrix
is denoted by $\sigma_0 = \mathds{1}$ in this manuscript. These $\sigma$-matrices describe the degree of freedom of a lattice pseudo-spin.
The Hamiltonian describes a massless Weyl fermion. The correspondent spectrum is :
\begin{equation}
E_{\lambda}(q_x,q_y) =  \lambda v_F\, \sqrt{q_x^2 + q_y^2} \; ,
\end{equation}
in which $\lambda =\pm \, 1$ defines the conduction and valence bands, respectively. The Hamiltonian \eqref{H1} commutes with the chirality operator
\begin{equation}
\mathcal{\eta} = \frac{ q_x \, \sigma^x +  q_y \, \sigma^y}{\sqrt{ q_x^2 +q_y^2}} \; ,
\end{equation}
whose the eigenvalues read as $\alpha = \pm \, 1$. Due to the honeycomb structure of the graphene lattice, there is another two-fold degeneracy of the Dirac cones $D$ and $D'$, which is set by the index $\rho = \pm \, 1$. Therefore, the band index can be properly identified as $\lambda = \rho \, \alpha$, see Fig. \eqref{fig1}. The "left" electrons ($\alpha= +1$) in the $D$ cone are in the conduction band, whereas the "right" electrons ($\alpha = -1$) are in the valence band (also in the $D$ cone), the inverse occurs in the $D'$ cone.
Including these degeneracies and assuming for simplicity $v_F=1$, we can write a 4-component Weyl spinor $(\psi)$ that satisfies the following massless Dirac Lagrangian :
\begin{equation}\label{lagr1}
\mathcal{L}= i \, \overline{\psi} \, \left[ \gamma^0 \, \partial_t - \gamma^x \, \partial_x - \gamma^y \, \partial_y \right]  \psi
= i \, \overline{\psi} \, \gamma^\mu \, \partial_\mu \psi \; ,
\end{equation}
where the $4 \times 4$ $\gamma^{\mu}$-matrices are defined by
\begin{equation}
\gamma^\mu =  \begin{pmatrix}
1 & 0 \\
0 & - 1
\end{pmatrix} \otimes \sigma^\mu  =  \begin{pmatrix}
\sigma^\mu & 0 \\
0 & -\sigma^\mu
\end{pmatrix}
\; ,
\end{equation}
with $\sigma^{\mu} = (\sigma^3, i \sigma^x, i\sigma^y)$, and $\overline{\psi} = \psi^\dagger \gamma^0$ is the adjoint spinor. The $\gamma^{\mu}$-matrices satisfy the Clifford algebra $\{\gamma^\mu, \gamma^\nu\}= 2 \eta^{\mu \nu} \mathds{1}$, and the identity $\gamma^\mu \gamma^\nu = \eta^{\mu \nu} \mathds{1}  + i\epsilon^{\mu \nu \lambda} \gamma_3 \gamma_\lambda$, where $\mathds{1}= \mathds{1}_{4 \times 4}$,
$\gamma_3 = \begin{pmatrix}
\mathds{1}_{2 \times 2} & 0\\
0 & -\mathds{1}_{2 \times 2}
\end{pmatrix}$
(for details see the ref. \cite{Rosenstein91}). The non-trivial traces involving these $\gamma^{\mu}$-matrices are :
\begin{subequations}
\begin{eqnarray}
\mbox{Tr} \left( \gamma^\mu \gamma^\nu \right) \!\!&=&\!\! 4 \eta^{\mu \nu} \; , \;
\mbox{Tr} \left( \gamma^\mu \gamma^\nu \gamma^\lambda\gamma_3 \right) =4 i \epsilon^{\mu \nu \lambda} \, ,
\hspace{0.5cm}
\\
\mbox{Tr}\left( \gamma^\mu \gamma^\nu \gamma^\alpha \gamma^\beta\right) \!\!&=&\!\! 4 \left(\eta^{\mu \nu} \eta^{\alpha \beta} + \eta^{\mu \alpha} \eta^{\nu \beta} - \eta^{\mu \beta} \eta^{\nu \alpha} \right) \, . \;\;\;\;
\end{eqnarray}
\end{subequations}
The Dirac matrices also satisfy the identities :
\begin{subequations}
\begin{eqnarray}
\gamma_\mu \gamma^\mu \!&=&\! 3 \mathds{1} \; , \;
\gamma^\mu \gamma_\nu \gamma_\mu = - \, \gamma_\nu \; ,
\\
\gamma^\mu \gamma_\nu \gamma_\lambda \gamma_\mu \!&=&\! 3 \, \eta_{\nu \lambda} \mathds{1}
- i \,  \, \epsilon_{\nu \lambda \kappa} \,  \gamma^\kappa \, \gamma_3  \; ,
\\
\epsilon^{\mu \nu \kappa} \gamma_ \mu \gamma_\nu \!&=&\! -2 \, i  \, \gamma^\kappa \, \gamma_3 \; ,
\\
\epsilon^{\mu \nu \kappa} \gamma_ \mu \, \slashed{a} \, \gamma_\nu \!&=&\! -2 \, i \, a^\kappa \, \gamma_3 \; .
\end{eqnarray}
\end{subequations}
Note that we do not introduce the two-fold degeneracy of the true spin, and in the case of our interest, we have to double the spinors
as $\psi \rightarrow \psi_s$, for $s=\pm \, 1$. It is simple to check that the Lagrangian \eqref{lagr1} has the chiral symmetry
$\psi \rightarrow \gamma_5 \psi$, and $\overline{\psi} \rightarrow - \gamma_5 \overline{\psi}$, with $i\gamma_5 = \begin{pmatrix}
  0 & \mathds{1}_{2 \times 2} \\
  -\mathds{1}_{2 \times 2} & 0
  \end{pmatrix}$.
There are two possibilities of mass terms for the Lagrangian \eqref{lagr1}. We are able to write the following two bilinears, such that, $\overline{\psi} \, \psi$ and $\overline{\psi} \, \gamma_3 \, \psi$ for the like-massive terms. Both these massive bilinears break the chiral symmetry, but the $\gamma_3$-term also breaks the pseudo-spin degeneracy, the second term is commonly discarded. Nonetheless, both mass terms open a gap between the conduction and valence bands, and it is important to the study of the metal-insulator properties applied to the $2D$ materials. In this work, we will study both terms and we show that the $\gamma_3$-term generates new results in comparison with the standard massive term.
Moreover, the construction of a bilayer of graphene (BLG) can be achieved by doubling the degrees of freedom and introducing the proper coupling between the top and bottom electrons (see refs. \cite{McCann}). In the sequel, we discuss the BLG electronic low-energy model and analyze the results.

\section{ The bilayer graphene}	
\label{sec3}
\begin{figure}[htb!]
\centering
\includegraphics[scale=0.63]{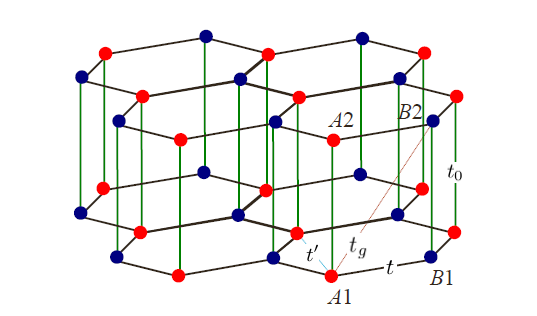}
\caption{The AA-type displacement. The parameters $t'$ and $t_g$ are one order of magnitude lesser than $t$ and $t_0$.}
\label{fig2}
\end{figure}

The structure formed by two or more layers of graphene was first reported in 2004 \cite{novoselov}.
There are two main kinds of graphene bilayers, the AA and the AB configurations \cite{Rozhkov16}. In the AA configuration, the atoms in the top layers are exactly above the bottom atoms, see Fig. \eqref{fig2}. In the AB configuration, the A-type sub-lattice atoms of the top layer are located above the B-type sub-lattice atoms of the bottom layer. Both cases are illustrated in ref. \cite{Rozhkov16}. Due to this configuration, the B-type sub-lattice atoms of the top layer (B2) are located above the empty center of the lower hexagon. This configuration modifies the dynamic of the quasi-particles, and the low energy of the AA type presents a linear energy spectrum, whereas the AB type presents a quadratic energy spectrum \cite{McCann}. Although less stable than the AB-type \cite{mostaani}, the AA-type is the only one that has the structure that maintains the Dirac-like low-energy Hamiltonian and because of this property will be the target of this work.
%

%
From the tight-binding description of the system, the energy eigenstates of the quasi-particles
in the AA configuration is given by \cite{Rozhkov16}:
\begin{equation}
\epsilon_{\ell,\lambda}({\bf q}) = \ell \, t_0 + \lambda \, v_F \, |{\bf q}|+ O(|{\bf q}|^2) \; ,
\end{equation}
where $\ell =+ 1$ for the top layer, $\ell=-1$ for the bottom layer, and $\lambda = \pm \, 1$ for electron/hole, as the usual.
The $t_{0}$-parameter has energy dimension, and it is estimated in the range of $0.3-0.4$ eV. It is called {\it interplane
nearest-neighbor hopping integral}, and is related to the interaction between the layers \cite{Gusynin}.
Going further, fixing $v_F=1$ for simplicity, the Lagrangian for the AA configuration can be written as :
\begin{eqnarray}
\label{lagrAABLGt}
\mathcal{L}^{AA} \!&=&\! \sum_{I=-1}^{+1} \, \overline{\psi}_I \left[ \, \gamma^0 \left(i \partial_t - I t_0\right)- i\,{\bf \gamma} \cdot {\bf \nabla} \, \right] \psi_I
\nonumber \\
\!&=&\! \sum_{I=-1}^{+1} \left( \, \overline{\psi}_I \, i \, \gamma^\mu \partial_\mu \psi_{I}  + I \, t_0 \, \overline{\psi}_I \, \gamma^0 \, \psi_I \, \right)
\nonumber \\
\!&=&\! \overline{\Psi} \, i \, \Gamma^\mu \, \partial_\mu \Psi + \overline{\Psi} \, \Lambda^\mu \, t_\mu \,  \Psi \; ,
\end{eqnarray}
where we have defined the $8$-component spinor $\Psi = \left( \, \psi_{+} \;\; \psi_{-} \, \right)^{t}$, and the time-like $(1+2)$-vector $t^\mu = (t_0,{\bf 0})$. In this stage, it is also convenient to define the $8 \times 8$ matrices :
\begin{subequations}
\begin{eqnarray}
\Gamma^{\mu} \!&=&\! \begin{pmatrix}
\gamma^{\mu} & 0 \\
0 & \gamma^{\mu}
\end{pmatrix}
\; , \;
\Gamma_{3} = \begin{pmatrix}
\gamma_{3} & 0 \\
0 & \gamma_{3}
\end{pmatrix}
\; , \;
\\
\Lambda^{\mu} \!&=&\! \begin{pmatrix}
\gamma^{\mu} & 0 \\
0 & -\gamma^{\mu}
\end{pmatrix}
\; , \;
\Lambda_{3} = \begin{pmatrix}
\mathds{1} & 0 \\
0 & -\mathds{1}
\end{pmatrix}
\; ,
\end{eqnarray}
\end{subequations}
%
%
that satisfy the Clifford algebra
\begin{subequations}
\begin{eqnarray}
\left\{ \, \Gamma^{\mu} \, , \, \Gamma^{\nu} \, \right\}
\!&=&\! 2\,\eta^{\mu\nu}\,\mathds{1}_{8}
\; , \;
\\
\left\{ \, \Lambda^{\mu} \, , \, \Lambda^{\nu} \, \right\}
\!&=&\! 2\,\eta^{\mu\nu}\,\mathds{1}_{8} \; ,
\\
\left\{ \, \Gamma^{\mu} \, , \, \Lambda^{\nu} \, \right\}
\!&=&\! 2\,\eta^{\mu\nu}\,\Lambda_3
\; , \;
\end{eqnarray}
\end{subequations}
and the commutator algebra
\begin{subequations}
\begin{eqnarray}
\left[ \, \Gamma^{\mu} \, , \, \Gamma^{\nu} \, \right] \!&=&\! \left[ \, \Lambda^{\mu} \, , \, \Lambda^{\nu} \, \right]= 2 \, i \, \Sigma^{\mu\nu} \; ,
\\
\left[ \, \Gamma^{\mu} \, , \, \Gamma_3 \, \right] \!&=&\! \left[ \, \Gamma^{\mu} \, , \, \Lambda_3 \, \right]=0 \, ,
\end{eqnarray}
\end{subequations}
where $\Sigma^{\mu\nu} =
   \epsilon^{\mu \nu \lambda} \, \Gamma_3 \, \Gamma_\lambda $.
These matrices also satisfy the hermitian properties :
$(\Gamma^{\mu})^{\dagger}=\Gamma^{0} \, \Gamma^{\mu} \, \Gamma^{0}$,
$(\Lambda^{\mu})^{\dagger}=\Gamma^{0} \, \Lambda^{\mu} \, \Gamma^{0}$, $(\Gamma_{3})^{\dagger}=\Gamma_{3}$
and $(\Lambda_{3})^{\dagger}=\Lambda_{3}$. In (\ref{lagrAABLGt}), the adjoint field is $\overline{\Psi}=\Psi^{\dagger} \, \Gamma^{0}$.
Thereby, the model \eqref{lagrAABLGt} is interpreted as a $(1+2)$-$D$
model summed to kinetic term that breaks the Lorentz symmetry through the time-like $(1+2)$-vector $t^\mu$,
analogous to the approach of a $(1+3)$-$D$ with a CPT-odd parameter $(b^\mu)$, see \cite{perez}.
Admittedly, the massless character of the electronic quasi-particles in the graphene is a well-studied problem, and in general, the chiral symmetry is broken by the presence of phonons, and for a regime of low temperature. Thus, we consider that the mass gap exists, and we introduce the most general mass term adding $\overline{\Psi} \, \Psi $ and $\overline{\Psi} \, \Gamma_3 \, \Psi$ in the lagrangian (\ref{lagrAABLGt}) :
\begin{equation}\label{Lfermionmass}
\mathcal{L}_{AA} = \overline{\Psi} \, i \, \Gamma^\mu \, \partial_\mu \Psi-m \, \overline{\Psi} \, \Psi-\mu \, \overline{\Psi} \, \Gamma_3 \, \Psi
+ \overline{\Psi} \, \Lambda^{\mu} \, t_{\mu} \, \Psi \; ,
\end{equation}
where $m$ and $\mu$ are real parameters with the mass dimension. The chiral symmetry breaking in Kekulé-ordered graphene estimates the quasi-particle mass at $m=0.19-0.21$ eV \cite{Bao}. In this sector, we observe that the massive eigenstates for the fermions are :
$M_{\pm}=m \, \pm \, \mu$, with an intra-layer mass gap of $\Delta M=M_{+}-M_{-}=2\mu$. From the lagrangian (\ref{lagrAABLGt}), the action principle
yields the field equations
\begin{subequations}
\begin{eqnarray}
i \, \Gamma^\mu \partial_\mu \Psi-m\,\Psi-\mu\,\Gamma_3\,\Psi + \Lambda^{\mu} \, t_\mu \, \Psi = 0 \; ,
\label{EqDiracmasspsi}
\\
i\,(\partial_\mu\overline{\Psi}) \, \Gamma^\mu + \overline{\Psi} \, m
+ \mu \, \overline{\Psi} \, \Gamma_3 - \overline{\Psi} \, \Lambda^\mu \, t_\mu = 0 \; ,
\label{EqDiracmasspsibar}
\end{eqnarray}
\end{subequations}
which when combined, lead to the continuity equation $\partial_{\mu}J^{\mu}=0$, where the conserved current is
$J^{\mu}=\overline{\Psi} \, \Gamma^\mu \, \Psi$. Using the plane wave solution $\Psi(x)=u(p) \, e^{-i \, p \, \cdot \, x}$,
the equations (\ref{EqDiracmasspsi}) and (\ref{EqDiracmasspsibar}) in the momentum space are read
\begin{subequations}
\begin{eqnarray}
\left( \, \Gamma^\mu\,p_\mu -m \, \mathds{1}_{8}-\mu\,\Gamma_3 + \Lambda^{\mu} \, t_\mu \, \right) u(p) = 0 \; ,
\label{EqDiracmasspsip}
\\
\overline{u}(p)\left( \, \Gamma^\mu \, p_{\mu} - m \, \mathds{1}_{8}
- \mu \, \Gamma_3 + \Lambda^\mu \, t_\mu \, \right) = 0 \; ,
\label{EqDiracmasspsibarp}
\end{eqnarray}
\end{subequations}
where $u(p)$ is the amplitude matrix (column matrix of 8 components), and $p^{\mu}$ is the wave-momentum in the $(1+2)$-space.
The combination of these two equations is as follow : multiplying (\ref{EqDiracmasspsip}) to the left by
$\overline{u}(p^{\prime}) \, \Gamma^{\mu}$, also the eq. (\ref{EqDiracmasspsibarp}) to the right by
$\Gamma^{\mu} \, u(p)$, and using the matrices properties, we obtain the Gordon identity
\begin{eqnarray}\label{Ju}
\overline{u}(p^{\prime}) \, \Gamma^{\mu} \, u(p)
=\overline{u}(p^{\prime})
\left[ \, \frac{p^{\mu}+p^{\prime\mu}}{2m}+\frac{i}{2m} \, \Sigma^{\mu\nu}\,q_{\nu}
\right.
\nonumber \\
\left.
+\frac{t^{\mu}}{2m}\,\Lambda_3
-\frac{\mu}{m} \, \Gamma^{\mu}\,\Gamma_{3} \, \right]u(p) \; ,
\end{eqnarray}
where $q_{\nu}=p_{\nu}-p_{\nu}^{\prime}$ is known as the electron's recoil momentum.
It is important to remark that both $t^{\mu}$-vector and the gap contribute to the current.
This result helps us to understand how the fermion spin couples to the EM field through
the term with $\Sigma^{\mu\nu}$. We will investigate it in the section \ref{sec4}.
\section{The AA configuration coupled to the EM field}
\label{sec4}

From the usual and well known $U(1)$ gauge symmetry principle, we couple the fermions from (\ref{Lfermionmass}) to the EM field through the covariant derivative operator $D_{\mu}=\partial_\mu + i \, e \, A_\mu$, {\it i.e.},
\begin{equation}\label{lagrAABLG}
\mathcal{L}_{AA} = \overline{\Psi} \, i \, \Gamma^\mu D_\mu \Psi-m \, \overline{\Psi} \, \Psi-\mu \, \overline{\Psi} \, \Gamma_3 \, \Psi
+ \overline{\Psi} \, \Lambda^{\mu} \, t_{\mu} \, \Psi \; ,
\end{equation}
where $e$ is the dimensionless coupling constant (electron's fundamental charge), the gauge field $A^{\mu}=(A^{0},A_{x},A_{y})$ is the vector potential of the electrodynamics in $(1+2)$-dimensions. The interaction reproduced here is similar to the quantum electrodynamics (QED), that is,
$-e\,\overline{\Psi} \, \Gamma^{\mu} A_{\mu} \, \Psi$ in $(1+2)$ dimensions.
In the momentum space, the free fermion propagator comes from the matrix
\begin{eqnarray}\label{S0inv}
S_0^{-1}(p) = \Gamma^{\mu}\,p_{\mu} + \Lambda^{\mu} \, t_{\mu}  - m \, \mathds{1}_{8} - \mu\,\Gamma_3 \; ,
\end{eqnarray}
whose inverse is given by
\begin{widetext}
\begin{eqnarray}\label{S0prop}
S_0(p) \!&=&\!\frac{\left(  \slashed{p} + \slashed{t} \Lambda_3 + m \mathds{1}_{8} + \mu \Gamma_3\right) \left[ \Xi^2- 2 (p \cdot t)\Lambda_3 +2 m \mu \Gamma_3\right]\left[\Xi^4 - 4 (p \cdot t)^2 -4 m^2 \mu^2 - 8 m \mu \, (p\cdot t) \, \Lambda_3 \, \Gamma_3 \right] }{\left[\Xi^4 - 4(p \cdot t)^2 -4 m^2 \mu^2\right]^2 - 64 \, m^2 \mu^2 \, (p \cdot t )^2}
 \; ,
\end{eqnarray}
\end{widetext}
%
%
where $\Xi^2 = p^2 +t^2 - m^2 - \mu^2$. The pole of the propagator emerges from the dispersion relation
\begin{equation}
\left[ \left( \Xi^2 \right)^2 \! - 4(p \cdot t)^2 -4 m^2 \mu^2\right]^2
\!- 64 \, m^2 \mu^2 \, (p \cdot t )^2 = 0 \; ,
\end{equation}
whose the solutions yield the energy as a function of the spatial linear momentum $({\bf p})$
\begin{subequations}
\begin{eqnarray}
(p_0)_{I,n} =- I \, t_0 \pm \sqrt{ { \bf p}^2+(m+n \mu )^2} \; ,
\end{eqnarray}
\end{subequations}
%
%
where $I = \pm 1$ is the inter-layer index, and $n = \pm 1$ is the intra-layer index. The fermion propagator (\ref{S0prop}) is equivalent to expression
\begin{eqnarray}\label{S0}
S_{0}(p)=\sum_{n=-1}^{+1} \sum_{I=-1}^{+1} \mathcal{P}_I \otimes  P_n \, \frac{ \slashed{p} +  I \, \slashed{t}+M_n}{ \left(p+I t\right)^2-M_n ^2} \; ,
\end{eqnarray}
where we have defined the mass eigenvalues $M_n:=m+n \, \mu$, and we write it in terms of the intra- and inter-layer projectors are, respectively, defined by
\begin{eqnarray}
P_{n}= \frac{\mathds{1}+ n \, \gamma_3}{2}
\hspace{0.3cm} \mbox{and} \hspace{0.3cm}
\mathcal{P}_I=\frac{\mathds{1}_{2 \times 2}+I \,\tau_3}{2} \; .
\end{eqnarray}
%
%
The dynamics of the EM sector are governed by the pseudo-electrodynamics lagrangian \cite{dudal}
\begin{equation}\label{Lgauge3D}
{\cal L}_{gauge}^{3D}=-\,\frac{1}{4 } \, F_{\mu\nu} \frac{2}{\sqrt{-\square}}\,F^{\mu \nu}
-\frac{1}{2\xi}\left(\partial_{\mu}A^{\mu}\right)\frac{1}{\sqrt{-\square}}\left(\partial_{\nu}A^{\nu}\right) \; ,
\end{equation}
where $F_{\mu\nu}=\partial_{\mu}A_{\nu}-\partial_{\nu}A_{\mu}$ is the EM field strength tensor,
and $\xi$ is a gauge fixing parameter. In $(1+2)$-dimensions, the strength field tensor has the components
$F^{\mu\nu}=\left(E^{i},B\right) \, (i=1,2)$, in which the electric field acts on ${\cal XY}$-plane,
and the magnetic field become a pseudo-scalar in planar systems.
The contraction of the current (\ref{Ju}) with the
$A^{\mu}$-potential yields the pseudo-spin Hamiltonian coupled to the EM field  :
\begin{eqnarray}
H_{ps}=\frac{ie}{2m} \, \Sigma_{\mu\nu}\,q^{\nu}\,A^{\mu}=\frac{\mu_{B}}{2} \, \Sigma_{\mu\nu}F^{\mu\nu} \; ,
\end{eqnarray}
where we have used $q_{\nu} \rightarrow i\partial_{\nu}$, and $\mu_{B}=e/(2m)$ is the Bohr's magneton of the quasi-particle.
Using the definition of $\Sigma^{\mu\nu}$, the pseudo-spin Hamiltonian can be written as
\begin{eqnarray}\label{Hspin}
H_{ps}=-\frac{\mu_{B}}{4} \, \epsilon^{\mu\nu\lambda}\,F_{\mu\nu}\,J_{3\lambda} \; ,
\end{eqnarray}
in which $J_{3\lambda}=\overline{u}(p^{\prime}) \, \Gamma_{\lambda} \, \Gamma_{3} \, u(p)$.

This result shows the contribution from the pseudo-spin for the magnetic dipole momentum is equivalent to the coupling of the "chiral" charge density $\rho_3$ with an external magnetic field $B$.

Since the coupling of the interaction of the fermions with the gauge field is small, we use the usual formalism from QFT
to calculate the perturbative contributions at the one-loop approximation. The quadratic effective action
associated with the gauge lagrangian (\ref{Lgauge3D}) in the momentum space is :
\begin{eqnarray}\label{Seff2k}
S_{eff}^{(2)}(\tilde{A}) = - \int \frac{d^3k}{(2\pi)^3} \, \tilde{A}^{\mu}(-k) \times
\nonumber \\
\times \left[ \,\eta_{\mu\nu}\,\sqrt{k^2}-\left(1-\frac{1}{2\xi}\right)\frac{k_{\mu}\,k_{\nu}}{\sqrt{k^2}}-\Pi_{\mu\nu}(k) \, \right] \tilde{A}^{\nu}(k) \; ,
\end{eqnarray}
where $\tilde{A}^{\mu}$ is the Fourier transform of $A^{\mu}$, and the fermion sector contributes for
the vacuum polarization tensor $\Pi_{\mu\nu}(k)$ at the order of $e^2$. Using the fermion propagation (\ref{S0}),
the vacuum polarization in this approximation is given by the traced integral
%
%
\begin{widetext}
\begin{eqnarray}\label{Pimunuint}
\Pi_{\mu \nu}(k) \!&=&\! - 2 e^2 \sum_{n=-1}^{+1}\sum_{n'=-1}^{+1} \int  \frac{d^3p}{(2 \pi)^3}
\frac{ \mbox{Tr} \Big[ \gamma_\mu P_n(\slashed{p}+\slashed{k}+M_n)\gamma_\nu P_{n'}(\slashed{p}+M_n) \Big]}{ \left[\,(p+k)^2 - M_n^2\,\right]\left( p^2-M_{n'}^2\right)}
\nonumber \\
&&
\hspace{-0.5cm}
=-2 e^2 \sum_{n=-1}^{+1} \int  \frac{d^3p}{(2 \pi)^3} \frac{ \mbox{Tr} \Big[ \gamma_\mu P_n (\slashed{p}+\slashed{k}+M_n)\gamma_\nu (\slashed{p}+M_n) \Big]}{ \left[\,(p+k)^2 - M_n^2\,\right]\left( p^2-M_n^2\right)} \; ,
\end{eqnarray}
\end{widetext}
where $k^{\mu}$ sets the photon external momentum, we have applied the shift $p \rightarrow p - I \, t$,
and the projector property $P_n \, P_{n'} = P_n \, \delta_{n n'}$. The polarization tensor
can be split in two terms :
\begin{equation}\label{Pimunu}
\Pi_{\mu \nu}(k) = \Pi_{\mu \nu}^{(s)}(k) + \epsilon_{\mu \nu \kappa} \Pi^\kappa(k) \; .
\end{equation}
The symmetric part is given by
\begin{eqnarray}
\Pi^{(s)}_{\mu \nu}(k) &=&  -8 e^2 \sum_{n=-1}^{+1}\int  \frac{d^3p}{(2 \pi)^3}\times
\nonumber \\
&&\hspace{-1.cm}
\frac{\left[ M_n^2 - p \cdot(p+k )\right]\eta_{ \mu \nu} + (p+k)_{\mu} \, p_{\nu} }{ \left[(p+k)^2 - M_n^2\right]\left( p^2-M_n^2\right)}=
\nonumber \\
&&\hspace{-1.cm}= I_0(k^2) \, \eta_{\mu \nu} + I_1(k^2) \, k_\mu \, k_\nu \; ,
\end{eqnarray}
where $I_{0}$ and $I_{1}$ are defined by
%
\begin{eqnarray}
I_0(k^2) \!\!&=&\!\! 4\alpha \, M_{+} \left[ \tanh^{-1}\left(\frac{\sqrt{k^2}}{2M_{+}}\right)
+g\left(\frac{k^2}{M_{+}^2}\right)
\right.
\nonumber \\
&&
\hspace{-1.0cm}
\left.
+ \frac{k^2}{M_{+}^2} \, f\left( \frac{k^2}{M_{+}^2}\right)\right]
\nonumber \\
&&
\hspace{-1.0cm}
+4\alpha M_{-}\left[\tanh^{-1}\left(\frac{\sqrt{k^2}}{2M_{-}}\right)
+g\left(\frac{k^2}{M_{-}^2}\right)
\right.
\nonumber \\
&&
\hspace{-1.0cm}
\left.
+ \frac{k^2}{M_{-}^2} \, f\left( \frac{k^2}{M_{-}^2}\right)
\right]
\, ,
\nonumber \\
I_1(k^2) \!\!&=&\!\! 2 \alpha \left[ \, \frac{1}{M_{+}} \, f\left( \frac{k^2}{M_{+}^2} \right)
+\frac{1}{M_{-}} \, f\left( \frac{k^2}{M_{-}^2} \right) \, \right] \, , \;\;
\end{eqnarray}
%
and $\alpha=e^2/(4\pi)=1/137$ is the fine structure constant. It is important
to highlight that in $D=3$, the pseudo-electrodynamics keeps the
coupling constant $(e)$ dimensionless. The gauge field $(A^{\mu})$ has a mass dimension,
and consequently, the EM field has a dimension of mass squared.
By convenience, we also have defined
the functions $f$ and $g$ in the appendix (see the formulas (\ref{fk}) and (\ref{gk})).
The anti-symmetric component of the polarization tensor
can be written as follows :
\begin{eqnarray}\label{Pimu}
\Pi^{\mu}(k) &=&  -8 \, i \, e^2 \, k^{\mu} \sum_{n=-1}^{+1} n \, M_n
\int  \frac{d^3p}{(2 \pi)^3}
\nonumber \\
&&
\hspace{-1.5cm}
\times \, \frac{1}{ \left[(p+k)^2 - M_n^2\right](p^2-M_n^2)}=
i \, I_{2}(k^2) \, k^{\mu} ,
\hspace{0.5cm}
\end{eqnarray}
in which the function $I_{2}(k^2)$ is
\begin{equation}\label{I2}
I_{2}(k^2)=4\alpha \left[ \coth^{-1}\left( \frac{\sqrt{k^2}}{2M_{+}} \right)- \coth^{-1}\left( \frac{\sqrt{k^2}}{2M_{-}} \right) \right] \; .
\end{equation}
%
Notice that $I_{2}(k^2)$ vanishes in the limit $\mu \rightarrow 0$, and consequently, the antisymmetric part of (\ref{Pimunu}) is null.
Substituting all these results in the effective action (\ref{Seff2k}), it can be written as
\begin{eqnarray}\label{Seff2kO}
S_{eff}^{(2)}(\tilde{A}) = - \int \frac{d^3k}{(2\pi)^3}
\, \tilde{A}^{\mu}(-k) \, {\cal O}_{\mu\nu}(k^2) \, \tilde{A}^{\nu}(k) \; ,
\end{eqnarray}
where the ${\cal O}_{\mu\nu}(k^2)$ matrix is
\begin{eqnarray}
{\cal O}_{\mu\nu}(k^2) \!&=&\! \left[ \, \sqrt{k^2}-I_{0}(k^2) \, \right] \theta_{\mu\nu}+
\nonumber \\
&&
\hspace{-1.7cm}
+\left[ \frac{\sqrt{k^2}}{2\xi}-I_{0}(k^2)-k^2\,I_{1}(k^2) \right]\rho_{\mu\nu}
-I_{2}(k^2) \, S_{\mu\nu} \, ,
\; \; \;
\end{eqnarray}
and the projectors in the momentum space are
\begin{eqnarray}
\theta_{\mu\nu}=\eta_{\mu\nu}-\rho_{\mu\nu}
\; , \;
\rho_{\mu\nu}=\frac{k_{\mu}\,k_{\nu}}{k^2}
\; , \;
S_{\mu\nu}=\epsilon_{\mu\nu\lambda}\,i\,k^{\lambda} \; .
\end{eqnarray}
The gauge propagator corrected to one loop is so obtained
by the inverse of ${\cal O}_{\mu\nu}(k^2)$, such that,
${\cal O}_{\mu\alpha}({\cal O}^{-1})^{\alpha\nu}=\delta_{\mu}^{\;\,\,\nu}$.
Using the properties of the projectors
\begin{eqnarray}
\theta^{\mu\alpha}\theta_{\alpha\nu} \!\!&=&\!\! \theta^{\mu}_{\;\;\,\nu}
\; , \;
\rho^{\mu\alpha}\rho_{\alpha\nu}=\rho^{\mu}_{\;\;\,\nu}
\; , \;
\theta^{\mu\alpha}\rho_{\alpha\nu}=0 \; ,
\nonumber \\
\theta^{\mu\alpha}S_{\alpha\nu} \!\!&=&\!\! S^{\mu}_{\;\;\,\nu}
\; , \;
\rho^{\mu\alpha}S_{\alpha\nu}=0
\; , \;
S^{\mu\alpha}S_{\alpha\nu}=-k^2\,\theta^{\mu}_{\;\;\,\nu} \; ,
\hspace{0.5cm}
\end{eqnarray}
we obtain
\begin{eqnarray}\label{propaAc}
\Delta_{\mu\nu}^{(c)}(k^2) \!&=&\! \frac{-i\,[\sqrt{k^2}- I_{0}(k^2)]}{[\sqrt{k^2}- I_{0}(k^2)]^2+k^2\,[I_{2}(k^2)]^2}\left( \eta_{\mu\nu}-\frac{k_{\mu}k_{\nu}}{k^2} \right)
\nonumber \\
&&
\hspace{-0.5cm}
-\frac{i \, 2\xi }{\sqrt{k^2}-2\xi\,I_{0}(k^2)-2\xi\,k^2\,I_{1}(k^2)}\frac{k_{\mu} k_{\nu}}{k^2}
\nonumber \\
&&
\hspace{-0.5cm}
+\frac{I_{2}(k^2) \, \epsilon_{\mu\nu\rho} \, k^{\rho}}{ [\sqrt{k^2}- I_{0}(k^2)]^2+k^2\,[I_{2}(k^2)]^2 } \; ,
\end{eqnarray}
in which the notation $(c)$ means the correction by the one loop integral (\ref{Pimunuint}).
This propagator has good behavior in the ultraviolet regime,
where all the terms go to zero when $k \rightarrow \infty$.
The last term in (\ref{propaAc}) is like a Chern-Symons propagator induced by the radiative correction of
$I_{2}(k^2)$. The propagator pole that previously was at $k^2=0$ in (\ref{Lgauge3D}),
now it is removed in the propagator (\ref{propaAc}) due to presence of the radiative corrections of $I_{0}(k^2)$ and $I_{2}(k^2)$.
In the perturbative formalism, we contract (\ref{propaAc}) with two classical conserved
currents that constraint the condition $k_{\mu}J^{\mu}=0$ in the momentum space :
\begin{eqnarray}
J^{\mu} \, \Delta^{(c)}_{\mu\nu}(k^2) \, J^{\nu}=\frac{- \, i \, J_{\mu}J^{\mu} \, [\sqrt{k^2}-I_{0}(k^2)] }{ [\sqrt{k^2}-I_{0}(k^2)]^2+k^2\,[I_{2}(k^2)]^2}
. \hspace{0.7cm}
\end{eqnarray}
Therefore, the gauge propagator has now the pole evaluated at
\begin{eqnarray}\label{Eqk2}
[\sqrt{k^2}- I_{0}(k^2)\,]^2+k^2\,[I_{2}(k^2)]^2=0 \; .
\end{eqnarray}
This equation is hard to obtain exactly the solution for $k^{2}$ as a function of the masses and of the $\alpha$-constant.
However, we can consider approximations that help to understand the contributions of the radiative corrections.
In the infrared regime, when $k^2$ is very small, $k^2\,[I_{2}(k^2)]^2 \approx 0$, and the equation (\ref{Eqk2})
is reduced to
\begin{eqnarray}
k^2 \simeq [I_{0}(0)]^2 \simeq  \left(\frac{8\alpha m}{3}\right)^2 \; ,
\end{eqnarray}
where we have taken the limit $k^2 \rightarrow 0$ in $I_{0}(k^2)$. Thereby, the finite value of
$I_{0}(k^2)$, when $k^2 \rightarrow 0$, contributes with a mass for the gauge field given by :
\begin{eqnarray}\label{phmass}
m_{a} = \frac{8\alpha m}{3} \; .
\end{eqnarray}

\begin{figure}[htb!]
\centering
\includegraphics[scale=0.40]{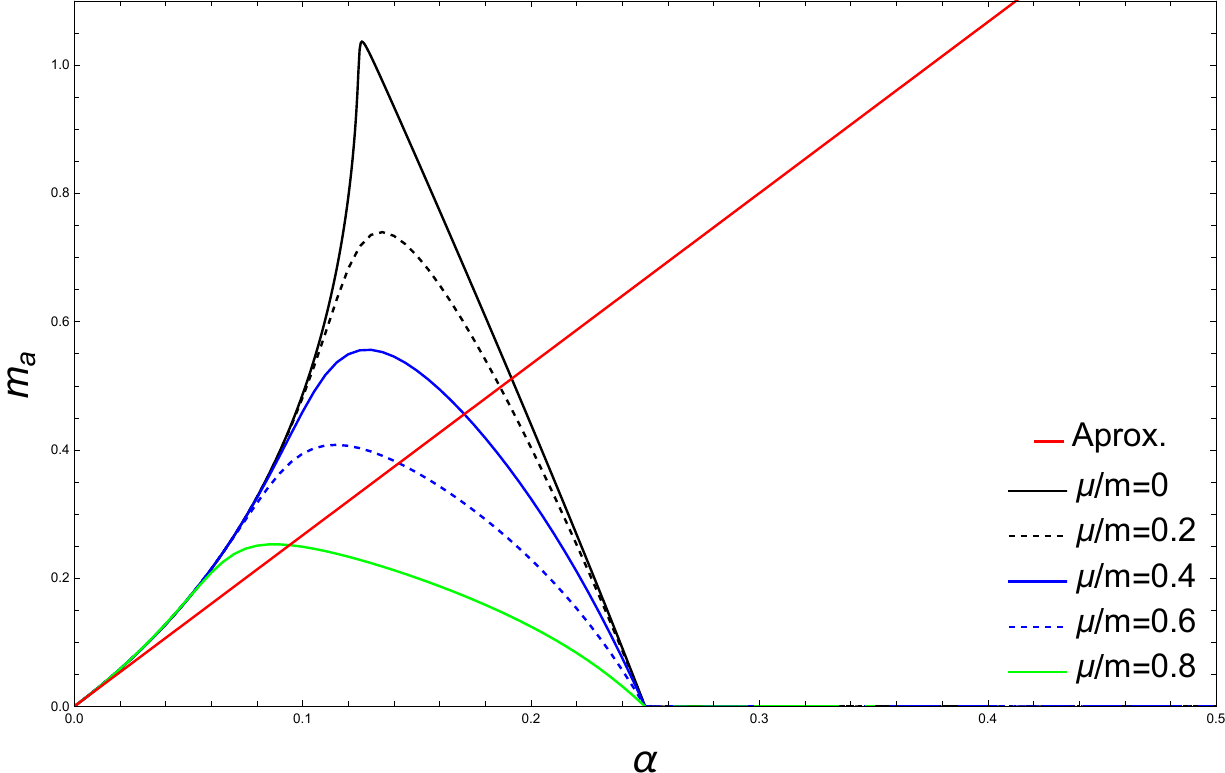}
\caption{The plot of the exact photon mass as a function of $\alpha$ for representative values of intra-layer gap $\mu$. The red line represents the approximation given by eq. \eqref{phmass}.}
\label{photonmass}
\end{figure}

By numerical calculations the values of the photon mass as a function of $\alpha$ can be found and it is shown in figure \ref{photonmass}. As can be seen, this approximation is valid only for $\alpha \ll 0.1$.

The action (\ref{Seff2kO}) in the coordinate space is the non-local Lagrangian
\begin{eqnarray}\label{LCFJ3DBox}
{\cal L}_{CFJ}^{3D} &\simeq& -\,\frac{1}{4} \, F_{\mu\nu} \frac{2}{\sqrt{-\Box}+\Box I_{1}(-\Box)}\, F^{\mu\nu}
\nonumber \\
&&
\!-\frac{1}{2\xi}(\partial_{\mu}A^{\mu})\frac{1}{\sqrt{-\Box}}(\partial_{\nu}A^{\nu})
\nonumber \\
&&
+\frac{1}{4} \, \epsilon^{\mu\nu\rho} \, A_{\mu} \, I_{2}(-\Box) \, F_{\nu\rho}
\nonumber \\
&&
+\frac{1}{2} \, A_{\mu} \left[ I_{0}(-\Box)-\Box\,I_{1}(-\Box) \right] A^{\mu} \; .
\end{eqnarray}
Still in the infrared regime, the effective Lagrangian from the action (\ref{LCFJ3DBox})
can be written as
\begin{eqnarray}\label{LCFJ3D}
{\cal L}_{CFJ}^{3D} \!&\simeq&\! -\,\frac{1}{4} \, F_{\mu\nu} \frac{2}{\sqrt{-\Box}}\, F^{\mu\nu}
\!-\frac{1}{2\xi}(\partial_{\mu}A^{\mu})\frac{1}{\sqrt{-\Box}}(\partial_{\nu}A^{\nu})
\nonumber \\
&&
\hspace{-0.7cm}
+\frac{\Delta}{4} \, \epsilon^{\mu\nu\rho} \, A_{\mu} \, \sqrt{-\Box}F_{\nu\rho}
-\frac{1}{2}\, \frac{8\alpha m}{3} \, A_{\mu}A^{\mu} \; ,
\hspace{0.5cm}
\end{eqnarray}
where the $\Delta$-parameter is
\begin{equation}
\Delta \simeq -\frac{4 \alpha  \mu }{m^2-\mu ^2} \; .
\end{equation}
Therefore, the corrections at one loop induce a massive Chern-Symons pseudo-electrodynamics in
$(1+2)$-dimensions when the low energy limit is applied to the gauge sector of the model.
This result shares some characteristics with the Proca Lagrangian of the pseudo-ED in ref. \cite{Ozela19},
although the Proca term presents a qualitative difference. 
\section{The fermion self-energy}
\label{sec5}
The self-energy of the fermionic quasi-particles can be calculated at one loop as follows :
\begin{eqnarray}
\Sigma(p) \!&=&\! -e^2\int \frac{d^3 k}{(2 \pi)^3} \, \Gamma^\mu \, S_0(p+k) \, \Gamma^\nu \, \Delta_{\mu \nu}(k)=
\nonumber
\\
&&
\hspace{-1.cm}
=-  e^2\sum_{n=-1}^{+1} \sum_{I=-1}^{+1}  \, \int \frac{d^3 k}{(2 \pi)^3} \frac{1}{\sqrt{k^2}}\frac{1}{ \left(p+k+I t\right)^2-M_n ^2} \;
\nonumber
\\
&&
\times \, \mathcal{P}_I \otimes  P_n \, \Gamma^\mu \left(\slashed{p} +\slashed{k}+  I \, \slashed{t}+M_n \right) \Gamma_\mu \; ,
\end{eqnarray}
where we use the free gauge propagator in the Feynman gauge $(\xi=1/2)$, {\it i.e.}, $\Delta_{\mu \nu}(k) = -i  \, \eta_{\mu \nu}/\sqrt{k^2}$.
This is the first correction to the fermion propagator due to the perturbative series that contributes to the mass of the quasi-particle.
The non-trivial contribution for the mass of the fermionic quasi-particle can be calculated via the application
of the trace over spinor space. The traces are read below :
\begin{subequations}
\begin{eqnarray}
\mbox{Tr}\left[\Sigma(p)\right] \!&=&\! \sum_{n=-1}^{+1} \sum_{I=-1}^{+1} \Xi_{I,n}(p) \, M_n \; ,
\\
\mbox{Tr}\left[\Gamma_3 \, \Sigma(p)\right] \!&=&\! \sum_{n=-1}^{+1} \sum_{I=-1}^{+1} \Xi_{I,n}(p) \, n \, M_n \; ,
\\
\mbox{Tr}\left[\Lambda_3 \, \Sigma(p)\right] \!&=&\! \sum_{n=-1}^{+1} \sum_{I=-1}^{+1}
\Xi_{I,n}(p) \, I \, M_n \; ,
\end{eqnarray}
\end{subequations}
in which $\Xi_{I,n}(p)$ is given by the integral
\begin{eqnarray}\label{Xinp}
\Xi_{I,n}(p)= -3 e^2\int \frac{d^3 k}{(2 \pi)^3} \frac{1}{\sqrt{k^2} }\frac{1}{ \left(k+p+I t\right)^2-M_n^2}
\; . \;\;\;\;
\end{eqnarray}
In the ultraviolet regime, this integral has a logarithmic divergence due to the
pseudo-electrodynamics propagator. Therefore, we introduce a dimensional regulator
parameter $(D)$ in which the original result is recovered in the limit $D \rightarrow 3$.
The regularized integral is :
\begin{eqnarray}\label{intSigma}
\Xi_{I,n}(p,D)
\!&=&\! - 3 e^2 (\Lambda)^{3-D} \times
\nonumber \\
&&
\times
\int \frac{d^D k}{(2 \pi)^D} \frac{1}{\sqrt{k^2}}\frac{1}{ \left(k+p+I t\right)^2-M_n ^2} \; ,
\nonumber\\
\end{eqnarray}
where $\Lambda$ is an arbitrary energy scale to keep the dimensionless coupling constant in $D$-dimension.
Using the technical shown in the appendix, we obtain
\begin{eqnarray}
\Xi_{I,n}(p,D) = \frac{3 \alpha}{16\pi} \, (-1)^{(D+3)/2}
\, \Gamma\left(\frac{3-D}{2}\right)
\times
\nonumber \\
\times
\int_0^1 dx \, (1-x)^{-1/2}
\left[\,\frac{4\pi\Lambda^2}{\Delta(x)^2}\,\right]^{(3-D)/2} \; ,
\end{eqnarray}
where $\Delta(x)^2=xM_{n}^2-x(1-x)(p+It)^2>0$, such that imposes the constraint of
$M_{n}^2>(p+It)^{2}$. It is evident that the previous expression is divergent if we make $D=3$ in the Gamma function.
To isolate the divergent term, we expand this result around $D=3$, with
$D=3-\delta$ taking the $\delta$-parameter very small $(\delta \rightarrow0)$. The result is reduced to $x$-integral
\begin{eqnarray}\label{Xidelta}
\Xi_{I,n}(p,\delta)=-\frac{3\alpha}{4\pi}\frac{1}{\delta}+\frac{3\alpha\gamma_{E}}{8\pi}
-\frac{3\alpha}{4\pi}-\frac{3\alpha}{8\pi} \ln\left(\frac{\pi\Lambda^2}{M_{n}^2}\right)
\nonumber \\
-\frac{3\alpha}{16\pi}\int_{0}^{1} \!\frac{dx} {\sqrt{1-x}}\ln\left[ \frac{M_{n}^2}{M_n^2 - (1-x)(p+ It)^2} \right] \; ,
\;\;\;
\end{eqnarray}
where $\gamma_{E}=0.577$ is the Euler-Mascheroni constant. The divergent term of (\ref{Xidelta}), $\delta^{-1}$ when $\delta \rightarrow 0$, and
the others one that does not depend on the external momentum are removed by a renormalization scheme of the model. Therefore, the finite part of
(\ref{Xidelta}) that depends on the external momentum $(p)$, and on the LSV parameter, gives the physical contribution that we are interested in.
Thereby, we obtain the result
\begin{eqnarray}\label{Xifinite1}
\Xi_{I,n}^{(f)}(p)
\!&=&\!
\frac{3\alpha}{8\pi}\left\{\frac{2M_{n}}{\sqrt{(p+It)^2}}\tanh^{-1}\left[ \frac{\sqrt{(p+It)^2}}{M_{n}} \right]
\right.
\nonumber \\
&&
\left.
-\ln\left[ \frac{M_{n}^2}{M_{n}^2-(p+It)^2} \right]
\right\}
\; .
\end{eqnarray}
 The limit of $p^{\mu}\rightarrow 0$ in (\ref{Xifinite1}) yields
\begin{equation}
\Xi_{n}^{(f)}(0)
\!=\!
\frac{3\alpha}{8\pi}\left[\frac{2M_{n}}{\sqrt{t^2}}\tanh^{-1}\left( \frac{\sqrt{t^2}}{M_{n}} \right)
-\ln\left( \frac{M_{n}^2}{M_{n}^2-t^2} \right)
\right] \; ,
\end{equation}
if $M_{n}^2>t^2$, and
\begin{eqnarray}
\Xi_{n}^{(f)}(0)
\!&=&\!
\frac{3\alpha}{8\pi}\left[\frac{2M_{n}}{\sqrt{t^2}}\tanh^{-1}\left( \frac{M_{n}}{\sqrt{t^2}} \right)
\right.
\nonumber \\
&&
\hspace{-0.5cm}
\left.
-\ln\left( \frac{M_{n}^2}{t^2-M_{n}^2} \right)-i\pi\left(1-\frac{M_{n}}{\sqrt{t^2}}\right)
\right] \; ,
\end{eqnarray}
if $t^2>M_{n}^2$.
Therefore, the corrections in the quasi-particle mass can be written as follows :
\begin{subequations}
\begin{eqnarray}
\delta m \!&=&\! \Re \left(\mbox{Tr}\left[\Xi(0)\right]\right) = 2 \sum_{n=-1}^{+1}  M_n \, \Re[\Xi_{n}(0)]
\nonumber
\\
&&
\hspace{-1.cm} \simeq \frac{6 \alpha m}{\pi} \left[-1+\ln\left(\frac{\sqrt{t^2}}{m}\right) \right]
\; ,
\label{deltam}
\\
\delta m_{IaL} \!&=&\! \Re\left( \mbox{Tr} \left[\Gamma_3 \, \Xi(0)\right]\right) = 2\sum_{n=-1}^{+1} \! n \, M_n \, \Re[\Xi_{n}(0)]
\nonumber \\
 &&
\hspace{-1.cm} \simeq \frac{6 \alpha \mu}{\pi}\left[-2+\ln\left( \frac{\sqrt{t^2}}{m} \right) \right]
\; ,
\\
\delta m_{IeL} \!\!&=&\!\! \mbox{Tr}\left[\Lambda_3 \, \Xi(0)\right] =
\!\sum_{I=-1}^{+1} \! I \! \sum_{n=-1}^{+1} \!\! M_n \, \Re[\Xi_{n}(0)] = 0
\; ,
\hspace{0.8cm}
\end{eqnarray}
\end{subequations}
where we have used the approximation $ m \gg \mu$ in these results. The inter-layer contribution $(\delta m_{IeL})$
to the quasi-particle mass is null in the limit of $p^{\mu} \rightarrow 0$, whereas the intra-layer $(\delta m_{IaL})$
contribution depends on the parameters $t^{\mu}$, $m$ and $\mu$. The self-energy has also an imaginary part and is given by:
\begin{equation}\label{se_im}
\gamma \approx \Im \left(\mbox{Tr}\left[\Xi(0)\right]\right) = \frac{3 \alpha m }{4}\left(1-\frac{m}{\sqrt{t^2}} \right) \Theta(\sqrt{t^2}-m) \; ,
\end{equation}
where $\Theta(x)=1$ for $x>0$ and $\Theta(x)=0$ otherwise, which means that quasi-particles with $ m < t_0 $ become unstable.

\section{The application in hydrodynamics}
\label{sec6}
In the limit with no collision, the kinetic equation for the fermionic system in a constant electromagnetic background field can be written as follows
\cite{hidaka} :
\begin{equation}
\left(  \, \slashed{p} + \frac{i \hbar}{2} \, \slashed{\nabla} -m \mathds{1}_8 - \mu \, \Gamma_3 +  \, \Lambda_3 \, \slashed{t} \, \right) F(x,p)  = 0 \; ,
\end{equation}
where the slashed operator $\slashed{\nabla}=\Gamma^{\mu}\nabla_{\mu}$ is defined by
%
$\nabla_\mu = \partial_\mu - e \, F_{\mu \nu} \, \partial^\nu_p$, in which $\partial^\nu_p$ means the derivative in relation to the momentum $p^{\mu}$,
%
and $F(x,p)$ is the Wigner function. The assumption of constant electromagnetic field implies $\partial_\lambda F_{\mu \nu} = 0$. It is convenient to decompose the Wigner function in terms of the Clifford algebra, using the definition of the projectors,
\begin{equation}
F(x,p) =\sum_{I=-1}^{+1} \sum_{n=-1}^{+1} \mathcal{P}_I \otimes P_n  \, F_{I,n}(x,p) \; ,
\end{equation}
in which $F_{I,n}$ satisfies the equation
\begin{equation}
\left(  \, \slashed{p} + \frac{i \hbar}{2} \, \slashed{\nabla} + I \, \slashed{t} - M_n \, \right) F_{I,n}(x,p)  = 0 \; .
\end{equation}
We write $F_{I,n}$ as $F_{I,n}= \slashed{\mathcal{V}}_{I,n} + \mathcal{F}_{I,n}$, and taking the trace, we obtain the set of equations :
\begin{subequations}
\begin{eqnarray}
\left( p_\mu  + I \, t_\mu \right) \mathcal{V}_{I,n}^\mu - M_n \, \mathcal{F}_{I,n} = 0 \; ,
\\
\nabla_\mu\mathcal{V}_{I,n}^\mu=0 \; ,
\\
\left( p_\mu  + I \, t_\mu \right) \mathcal{F}_{I,n} - M_n \,  (\mathcal{V}_{I,n})_\mu=0  \; .
\end{eqnarray}
\end{subequations}
Thus, we write
\begin{equation}
F_{I,n} = \frac{(\slashed{p} + I \, \slashed{t} + M_n )}{M_n} \, \mathcal{F}_{I,n} \; ,
\end{equation}
in which $\mathcal{F}_{I,n}$ satisfies the quadratic equation
\begin{equation}\label{sol1}
\left[ \, \left(p+ I \, t \right)^2 - M_n^2 \, \right]  \mathcal{F}_{I,n}=0 \; .
\end{equation}
Following the ref. \cite{Kondepudi}, one assumes the local equilibrium, {\it i. e.}, when the equilibrium thermodynamic relations are valid for the thermodynamic variables locally assigned. If this statement is valid, we can assume that the intensive thermodynamic variables are functions of the space-time coordinates. For our proposal, the temperature is $T = T(x)$ (for simplicity one fixes the same temperature in both layers), and $\alpha_I(x) = (\mu_{c})_I(x)/T(x)$ is the so-called fugacity,
where $(\mu_{c})_{I}(x)$ is the chemical potential. Therefore, the local Fermi-Dirac distribution can be written as $f_{FD}(x,p)= f_{FD}(\beta \cdot p - \alpha_I)$, with $f_{FD}(y) = (e^y + 1)^{-1}$, and $\beta^{\mu}_I(x) = u^{\mu}_I(x)/T(x)$, in which $u_I^{\mu}(x)$ is called local fluid velocity of the $I$-layer, such that it depends on $x$-coordinates only via the intensive thermodynamic variables. Finally, using eq. \eqref{sol1}, assuming a Fermi-Dirac characteristic of the system, and omitting the layer index $I$ for the sake of compactness, one can rewrite $\mathcal{F}_{I,n}$ as follows :
\begin{equation}
\mathcal{F}_{I,n}(x,p) = f_{FD}(\beta \cdot p - \alpha) \, \delta\left[ \, \left(p+ I t\right)^2 - M_n^2 \, \right] \; .
\end{equation}
Applying this solution in \eqref{sol1}, we obtain
\begin{equation}
\left[ \, p^\mu \, p^\nu \, \partial_\mu \beta_\nu
- p^\mu \left( \partial_\mu \alpha - I t^\nu \partial_\nu \beta_\mu +\beta^\nu F_{\mu \nu} \right) \right] =0 \; ,
\end{equation}
which implies, for $p^{\mu} \neq 0$, into the equations
\begin{subequations}
\begin{eqnarray}
\partial_\mu \beta_\nu + \partial_\nu \beta_\mu = 0 \; ,
\label{beta}
\\
\partial_\mu \alpha + I \, \epsilon_{\mu \nu \rho} \, t^\nu \, \omega^\rho + \beta^\nu \, F_{\mu \nu} = 0 \; ,
\label{newsol}
\end{eqnarray}
\end{subequations}
where $\omega^\rho = \epsilon^{\mu \nu \rho} \, \partial_\mu \beta_\nu $ is
the thermal vorticity vector, and $\beta=T^{-1}$ is the inverse of the temperature. The equations (\ref{beta}) and (\ref{newsol}) are
independent, and set the equilibrium conditions for a $(1+2)$-dimension system under the presence of a Lorentz violation
and constant electromagnetic field. The solution for eq. \eqref{newsol} is given by :
\begin{equation}
\beta_\mu (x) = (\beta_0)_\mu -\epsilon_{\mu \nu \lambda} \, x^\nu \, \omega_0^\lambda \; ,
\end{equation}
with $(\beta_0)_\mu$ and $\omega_0^\lambda$ constants, and
\begin{eqnarray}
\alpha_{I}(x) \!&=&\! \alpha_0 + \beta_0^\mu x^\nu F_{\mu \nu} - I \, \epsilon_{\mu \nu \rho} \, x^\mu \, t^\nu \, \omega_0^\rho
\nonumber \\
&&
+\frac{1}{2} \, \epsilon^{\nu \kappa \lambda} \, x^\mu \, x_\kappa \, (\omega_0)_{\lambda}F_{\nu \mu} \; ,
\end{eqnarray}
where $\alpha_{0}$ is an integration constant. The effect of the external electromagnetic field does not bring any new feature beyond the well-known Joule current, so one focuses on the LSV contribution. The solution for the fugacity $\alpha(x)$ implies the existence of a chemical affinity \cite{Kondepudi} in the $I$- layer given by $(A_I)_i \propto - \partial_i \alpha_I = - I t_0\epsilon_{ij } (\omega_0)_j$ (with $i = x,y$ the spacial coordinates) which means that, based on the Onsager reciprocal relations, it will generate a current perpendicular to the constant vortex vector $(\omega_0)_i $, will be proportional to the interplane neatest-neighbor hopping integral $t_0$ and with the signs $(+)$ and $(-)$ for the upper and lower layer, respectively.

\begin{figure*}[th]
\centering
\includegraphics[scale=0.43]{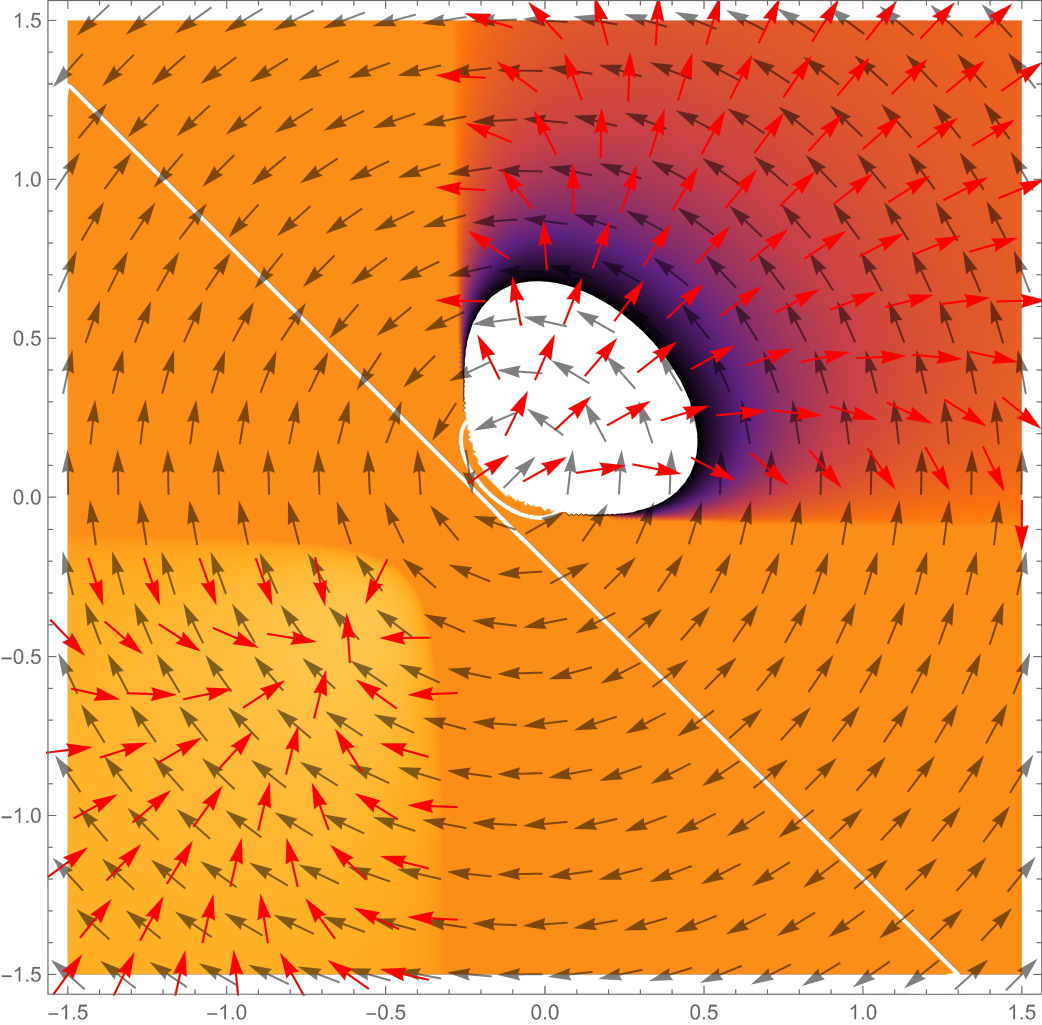}
\includegraphics[scale=0.43]{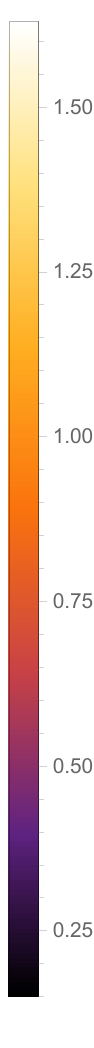}
\quad\quad
\includegraphics[scale=0.43]{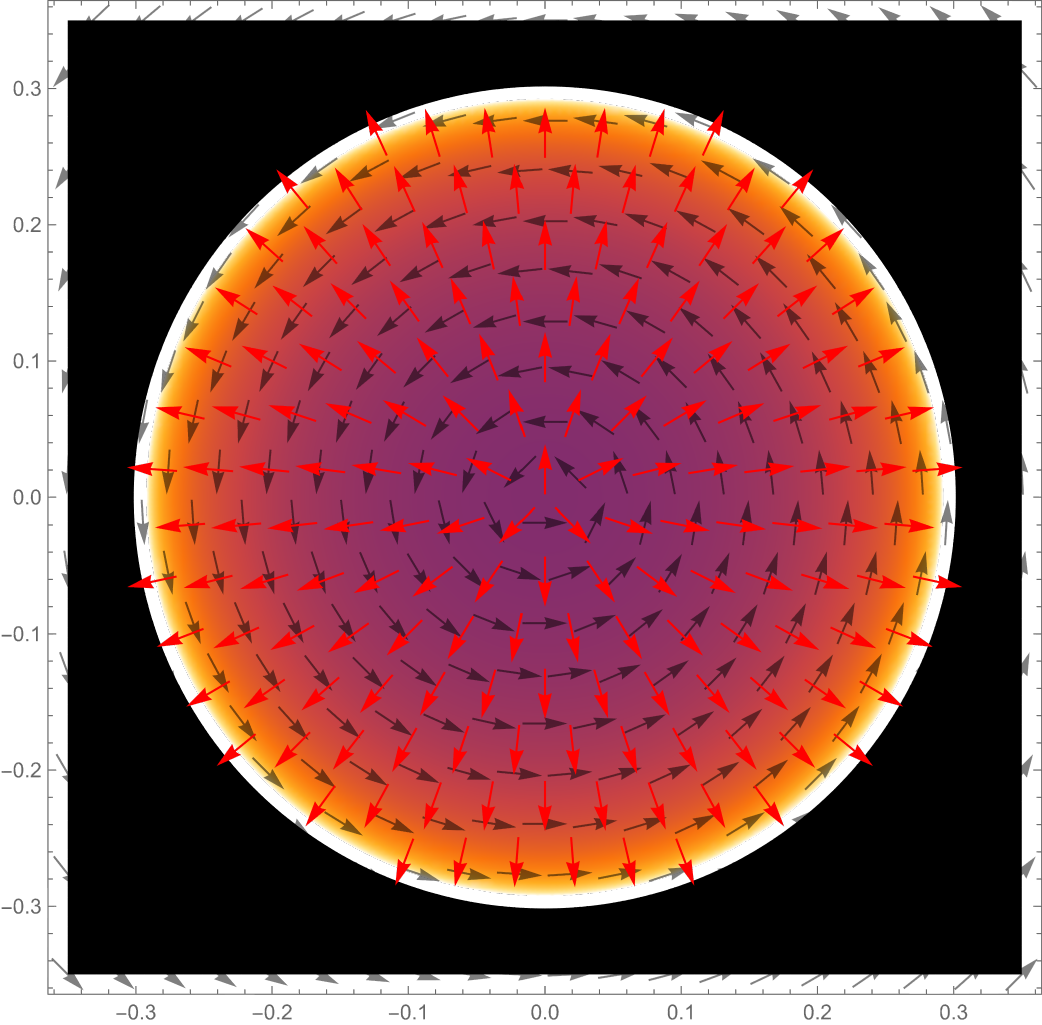}
\includegraphics[scale=0.43]{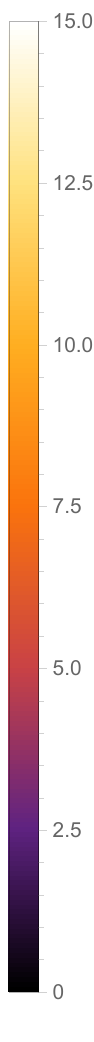}
\caption{Left panel : Plot of the temperature at $t=0$ in the presence of vortex with $\beta_0=(0.2,0,0)$ and $(\omega_0)=(0.1, 1,-1)$. The red arrow represents the temperature gradient, and the gray arrow represents the velocity flux. Right panel: Plot of the temperature at $t=0$ in the presence of vortex with  $\beta_0=(0.2,0,0)$ and $(\omega_0)=(0.1,0,0)$. The red arrow represents the temperature gradient, and the gray arrow represents the velocity flux.}
\label{vortex1}
\end{figure*}

To better visualize the effect, let's rewrite the fluid velocity in terms of the temporal and spatial components. Assuming the rest frame of the material, one defines $x^\mu=(t, x^i)$ and $u^\mu =\gamma (1, v^i)$, with $\gamma= (1-|v_i|^2)^{-1/2}$. After algebraic manipulations one finds the following steady-state solutions for $T(x^i,t)$ and $v^i(x^i,t)$ :
\begin{equation}\label{vprof}
v^i = \frac{(\beta_0)_i - \epsilon_{ij}\left[ \left(\omega_0\right)_0 x_j -t \left( \omega_0 \right)_j \right]}{(\beta_0)_0- \epsilon_{ij}x_i (\omega_0)_j} \; ,
\end{equation}
and
\begin{equation}
T(t,x^i) = (u \cdot \beta)^{-1} = \left( u \cdot \beta_0 - \epsilon_{\mu \nu \rho} \, u^\mu \, x^\nu \omega_0^\rho \right)^{-1} \; ,
\end{equation}
and can be checked that $v_i = (\beta_0)_i/(\beta_0)_0$ when $\omega_0 \rightarrow 0$ recovering the standard result of equilibrium hydrodynamics.  From eq. \eqref{vprof}, one can write the temperature in terms of $x^i$, and the integration constants $\beta_0$, $\omega_0$,
and the results are plotted in the fig. \ref{vortex1}.
Going further, the total current can be calculated and is given by :
\begin{eqnarray}
J^\mu(x) \!&=&\! \frac{1}{4}\int \frac{d^3p}{(2 \pi)^3} \, \mbox{Tr} \left[ \Gamma^\mu F(x,p) \right] =
\nonumber \\
&&\hspace{-1.cm} =   \sum_{I=-1}^{+1} \sum_{n=-1}^{+1}\int\frac{d^3p}{(2 \pi)^3} \, (p+ I t)^\mu \, f_{FD}(\beta \cdot p - \alpha_I) \, \times
\nonumber \\
&&
\times \, \delta\left[ \, \left(p+ I t\right)^2 - M_n^2 \, \right]
\nonumber \\
&&\hspace{-1.cm} =   \sum_{I=-1}^{+1} \sum_{n=-1}^{+1}\int\frac{d^3p}{(2 \pi)^3} \, p^\mu f_{FD}(\beta \cdot p - \tilde{\alpha}_I) \delta\left[ \,  p^2 - M_n^2 \, \right] \, ,
\nonumber \\
\end{eqnarray}
with $\tilde{\alpha}_I = \alpha_0 + I t \cdot \beta_0$. Therefore, temporal component of the current $(J^{0})$ is
\begin{eqnarray}
\rho(x)&=&\sum_{I=-1}^{+1} \sum_{n=-1}^{+1}\frac{T^2}{(2 \pi)^2} \times
\nonumber \\
&&
\hspace{-2.cm}
\times\int_{\beta |M_n|}^{\infty}dz  \left[ \frac{z}{e^{z-\tilde{\alpha}_I}+1}+  \frac{z}{e^{z+\tilde{\alpha}_I}+1} \right] \; .
\end{eqnarray}
The result of the integral is
\begin{eqnarray}
&&
\int_{\beta |M_n|}^{\infty}dz \left[ \frac{z}{e^{z-\tilde{\alpha}_I}+1}+  \frac{z}{e^{z+\tilde{\alpha}_I}+1} \right]
\nonumber
\\
&&=\beta \, |M_n| \ln \left[ \left(e^{-\tilde{\alpha}_I-\beta |M_n|}+1\right) \left(e^{\tilde{\alpha}_I-\beta  |M_n|}+1\right)\right]
+
\nonumber \\
&&-\text{Li}_2\left(-e^{-\tilde{\alpha}_I-\beta|M_n|  }\right)-\text{Li}_2\left(-e^{\tilde{\alpha}_I - \beta|M_n| }\right) \; ,
\end{eqnarray}
where $\mbox{Li}_{2}$ is a dilogarithm function. In the limit $t_0 >T \gg \left(\mu_I , M_n\right)$,
one reaches:
\begin{eqnarray}
\rho(x) \simeq \frac{t_{0}^2}{ 2 \pi^2}  \; .
\end{eqnarray}
The $\Lambda$-electronic current that measures the difference between the upper and lower layers is calculated by
\begin{eqnarray}
J_{\Lambda}^{\;\,\mu}(x) = \frac{1}{4}\int \frac{d^3p}{(2 \pi)^3} \, \mbox{Tr} \left[ \Gamma^\mu \Lambda_3 F(x,p) \right] \; .
\end{eqnarray}
In the limit of $t_0 >\mu_I \gg \left(T, M_n\right)$, one reaches:
\begin{eqnarray}
\rho_\Lambda (x) \simeq  \frac{t_0}{4 \pi^2} \left(\mu_+ + \mu_-\right) \; ,
\end{eqnarray}
where $\mu_\pm$ are the chemical potentials in the upper $(+)$ and lower $(-)$ layers, and implies that the combination of the LSV parameter with a non-null chemical potential induces a charge difference between the layers.

\section{Conclusions}
\label{sec7}
In this work, we study the low energy properties of the AA-type bilayer graphene in the context of Lorentz symmetry violation (LSV). We built up the fermionic sector of the quasi-particles of the model in which the structure of the material introduces naturally a time-like LSV parameter with energy dimension. This fermionic sector is coupled minimally to the electromagnetic field via the gauge principle. The dynamics of the EM field also is introduced, such that, an action of a quantum pseudo-electrodynamics (QED) in $1 + 2$ dimensions with LSV is so proposed. The properties from the point of view of a quantum field theory are studied in the paper. We calculate the contributions of the vacuum polarization and the self-energy for the fermionic quasi-particle. The vacuum polarization induces new terms in the effective lagrangian, which we interpret as the emergence of a massive term for the gauge field, and also of a Chern-Simons term in the low energy limit. These mass terms can be responsible to weaken the coulomb force and can facilitate the formation of chiral and superconducting gaps. On the other hand, from fig. \ref{photonmass} one can affirm that for $\alpha > 0.225$ the photon mass vanishes. This feature could be a limitation of the one-loop corrections and can change if we took contributions of higher loop contributions.

On the other hand, the fermion self-energy yields contributions to the quasi-particle mass due to perturbation theory inserted in the fermion propagator and can be used to seek information about the stability of the quasi-particles. Particularly, the imaginary part of the fermion self-energy given by \eqref{se_im} implies that the quasiparticles with effective mass $ m < t_0 \approx 0.3$ eV become unstable. These results point to a kind of protection against gap formation.
Important to highlight that we assume the gap formation when we introduce the massive terms. But, since the presence of the LSV parameter interferes with the formation of the gap and a study of the effects of LSV in the chiral symmetry breaking via Gross-Neveu models such refs. \cite{kneur1,kneur2,Gubaeva} can shed light on the limits of the presented model.

Additionally, we have shown the application of the fermionic system in the hydrodynamics approach, and show that the LSV parameter induces an anomalous current that couples with vortexes and generates a new kind of anomalous contribution to the hydrodynamic flux of the quasiparticles on the sheets of graphene. This new result is an anomalous thermal transport property of the BLG. The generation of a non-trivial fugacity, which is an explicit character of non-equilibrium systems can create new effects in the study of transport phenomena, in particular, the generation of anomalous currents via thermal gradients through and between the layers. Based on the fact that one assumes the no collision limit, the weakness of the Coulomb interaction becomes an implicit starting point. The study of the effects of the non-vanishing collision term can bring new information about the system beyond the small coupling regime.

Going further, a way to improve our result can be achieved by implementing high-order momentum corrections in the low-energy hamiltonian such as ref. \cite{pais}. One can also use the formalism presented in this work to describe the low-energy limit of the strained and twisted graphene bilayer, in which the constant vector $t^{\mu}$ becomes a space-dependent function related to the Moiré pattern formed by the strain/twist of the structure \cite{feng,sanjose,parhizkar}. Since the AA-type BLG has a small binding energy than the AB-type \cite{novoselov}, is natural to expect that curvature and torsion effects could affect the structure in a more intensive way, and our formalism can easily be generalized to accommodate these new phenomena. These features will be a target of analysis in forthcoming papers.
\section*{ACKNOWLEDGMENTS}
Y.M.P.G. is supported by a postdoctoral grant
from Funda\c c\~ao Carlos Chagas Filho de Amparo \`a Pesquisa do Estado do Rio de Janeiro (FAPERJ).
\appendix

\section{The Feynman integrals}
\label{appendix}

In this appendix, we show briefly the Feynman integral in one-loop that
were calculated in section IV. We start with the Feynman parametrization :
\begin{equation}
\frac{1}{AB} = \int^1_0 \frac{dx}{\left[A x + B(1-x) \right]^2} \; ,
\end{equation}
that allow us to join the propagator product
\begin{eqnarray}\label{intLmunu}
&& \frac{L_{\mu \nu}(p,q)}{ \left[(p+q)^2 - M_n^2\right]\left( p^2-M_n^2\right)}
\nonumber \\
&&=\int_0^1 \frac{dx \, L_{\mu \nu}(p,q)}{\left( 2 p \cdot q x + x q^2  + p^2 - M_n^2 \right)^2}
\nonumber \\
&&=\int_0^1 \frac{dx \, L_{\mu \nu}(p,q)}{\left[ (p+ x q)^2 +  x(1-x)q^2  - M_n^2 \right]^2}
\nonumber \\
&&=\int_0^1 dx \, \frac{L_{\mu \nu}(p-xq,q)}{\left(p^2  - \sigma^2 \right)^2} \; ,
\end{eqnarray}
where $L_{\mu \nu}(p,q)$ is a generic tensor that depends on the $p$ and $q$ momenta, and $\sigma^2 :=  M_n^2 - x(1-x) \, q^2$.
We have applied the shift $p \rightarrow p + x \, q$ in the last step of (\ref{intLmunu}). Using the known result from the
quantum field theory handbook \cite{Ryder}, the $D$-dimension integrals in the momentum space are :
%
%
\begin{eqnarray}
\int \frac{d^Dp}{(2\pi)^{D}}\frac{1}{(p^2 - \sigma^2)^\alpha} \!&=&\! \frac{\Gamma(\alpha - \frac{d}{2})}{(2\pi)^D\,\Gamma(\alpha)} \frac{(-1)^{\frac{D+1}{2}} \pi^{\frac{D}{2}} }{(- \sigma^2)^{\alpha - \frac{D}{2}}} \; ,
\\
\int \frac{d^Dp}{(2\pi)^{D}} \, \frac{p_\mu}{(p^2 - \sigma^2)^\alpha} \!&=&\! 0 \; ,
\\
\int \frac{d^Dp}{(2\pi)^{D}} \, \frac{p_\mu \, p_\nu}{(p^2 - \sigma^2)^\alpha} \!&=&\!
\frac{\Gamma(\alpha - 1-\frac{D}{2})}{(2\pi)^{D} \, \Gamma(\alpha)} \frac{(-1)^{\frac{D+1}{2}} \pi^{\frac{D}{2}} }{(- \sigma^2)^{\alpha -1- \frac{D}{2}}}
\, \eta_{\mu\nu} \; ,
\hspace{0.8cm}
\end{eqnarray}
where $\sigma^2 > 0$ is required in these results, and implies that $4M_{n}^2 > q^2$.
In $(1+2)$-dimensions, we make $D=3$ and $\alpha=2$, thus the previous results are reduced to
\begin{eqnarray}
\int \frac{d^3p}{(2\pi)^3} \frac{1}{(p^2 - \sigma^2)^2} \!&=&\! -\frac{i}{16\pi} \frac{1}{\sqrt{\sigma^2}} \; ,
\\
\int \frac{d^3p}{(2\pi)^3} \, \frac{p_\mu \, p_\nu}{(p^2 - \sigma^2)^2} \!&=&\! - \frac{i}{16\pi} \, \eta_{\mu \nu} \, \sqrt{\sigma^2} \; .
\end{eqnarray}
Thereby, the Feynman integral is
\begin{eqnarray}
&&\int  \frac{d^3p}{(2 \pi)^3} \frac{1}{ \left[(p+q)^2 - M_n^2\right]\left( p^2-M_n^2\right)}
\nonumber \\
&&= \int_0^1 dx \int \frac{d^3p}{(2 \pi)^3} \frac{1}{\left(p^2  - \sigma^2 \right)^2}
\nonumber \\
&& = \frac{ i }{8\pi} \int_0^1  \frac{dx }{\sqrt{ M_n^2 - x(1-x)\,q^2} }
\nonumber \\
&&  = -\frac{1 }{8 \pi |M_n|} \, \coth^{-1}\left( \frac{\sqrt{q^2}}{2M_{n}} \right) \; .
\end{eqnarray}
%
%
Going further, one has:
\begin{eqnarray}
&&\int \frac{d^3p}{(2\pi)^3} \frac{(p+q)_{\mu} \, p_\nu}{ \left[(p+q)^2 - M_n^2\right]\left( p^2-M_n^2\right)}
\nonumber \\
&&=\int_0^1 dx  \int \frac{d^3p}{(2\pi)^3} \frac{ p_\mu \, p_\nu + x (1-x) \, q_\mu \, q_\nu}{\left(p^2  - \sigma^2 \right)^2}
\nonumber \\
&&= -\frac{i |M_n|}{16 \pi} \left[ \, \eta_{\mu \nu} \, g(y)+\frac{ q_\mu \, q_\nu}{M_n^2} \, f(y) \, \right] \; ,
\end{eqnarray}
in which $y_{n} := q^2/M_{n}^2$, and the functions
$f$ and $g$ are defined by
\begin{eqnarray}\label{fk}
f(y_n) \!&=&\! \int_0^1 dx \, \frac{x(1-x)}{\sqrt{ 1-x(1-x) \,y_n} }
=
\nonumber \\
&&
\hspace{-1.cm}
= \frac{1}{4 y_n^{3/2}}\left[ \, -2 \sqrt{y_n}+(4-3y_n) \coth ^{-1}\left(\frac{2}{\sqrt{y_n}}\right)
\right.
\nonumber \\
&&
\left.
\hspace{-1.cm}
+4 y_n \mbox{csch}^{-1}\left(\sqrt{\frac{4}{y_n}-1}\right) \, \right] \; ,
\end{eqnarray}
and
\begin{eqnarray}\label{gk}
g(y_n) &=& \int_0^1 dx \sqrt{ 1 - x(1-x) \, y_n }=
\nonumber \\
&&
\hspace{-1.cm}
=\frac{2\sqrt{y_n}+(3y_n-4) \coth^{-1}\left(\frac{2}{\sqrt{y_n}}\right)}{4 y_n^{3/2}} \; .
\end{eqnarray}
Finally, the dimensionally regularized integral (\ref{Xinp}) that contributes to the self-energy at one loop is
\begin{eqnarray}\label{intSigma}
\Xi_{I,n}(p,D)
\!&=&\! - 3 e^2 (\Lambda)^{3-D} \times
\nonumber \\
&&
\times
\int \frac{d^D k}{(2 \pi)^D} \frac{1}{\sqrt{k^2}}\frac{1}{ \left(k+p+I t\right)^2-M_n ^2} \; ,
\nonumber\\
\end{eqnarray}
in which the result in (1+2) dimensions is recovered when $D=3$, and $\Lambda$ is an arbitrary energy
scale to keep the dimensionless coupling constant in $D$-dimensions. Using the Feynman parametrization
\begin{equation}
\frac{1}{A B^{1/2}} = \int_0^1 dx \, \frac{(1-x)^{-1/2}}{[\,x \, A + (1-x)\,B\,]^{3/2}} \; ,
\end{equation}
the integral (\ref{intSigma}) can be rewritten as :
\begin{eqnarray}
\Xi_{I,n}(p,D)
\!&=&\! - 3 e^2 (\Lambda)^{3-D} \int_0^1 dx \, (1-x)^{-1/2} \times
\nonumber \\
&&
\times
\int \frac{d^D k}{(2 \pi)^D} \frac{1}{\left[ \, k^2 - \Delta(x)^2 \, \right]^{3/2}} \; ,
\end{eqnarray}
where $[\Delta(x)]^2 = x M_n^2 - x(1-x)(p+ It)^2>0$, and imposes the condition $2M_{n}^2>(p+It)^2$.
By use of identity A3, we obtain :
\begin{eqnarray}
\Xi_{I,n}(p,D) = \frac{3 \alpha}{16\pi} \, (-1)^{(D+3)/2}
\, \Gamma\left(\frac{3-D}{2}\right)
\times
\nonumber \\
\times
\int_0^1 dx \, (1-x)^{-1/2}
\left[\,\frac{4\pi\Lambda^2}{\Delta(x)^2}\,\right]^{(3-D)/2} \; .
\end{eqnarray}
We expand around $D=3-\delta$, when $\delta \rightarrow 0$, the result is
\begin{eqnarray}\label{Xideltaap}
\Xi_{I,n}(p,\delta)=-\frac{3\alpha}{4\pi}\frac{1}{\delta}+\frac{3\alpha\gamma_{E}}{8\pi}
-\frac{3\alpha}{4\pi}-\frac{3\alpha}{8\pi}\ln\left( \frac{\pi\Lambda^2}{M_{n}^2} \right)
\nonumber \\
-\frac{3\alpha}{16\pi}\int_{0}^{1} \!\frac{dx} {\sqrt{1-x}}\ln\left[ \frac{M_{n}^2}{M_n^2 - (1-x)(p+ It)^2} \right] \, .
\;\;\;
\end{eqnarray}
The finite part is
\begin{eqnarray}\label{Xifinite1ap}
\Xi_{I,n}^{(f)}(p)
\!&=&\!
\frac{3\alpha}{8\pi}\left\{\frac{2M_{n}}{\sqrt{(p+It)^2}}\tanh^{-1}\left[ \frac{\sqrt{(p+It)^2}}{M_{n}} \right]
\right.
\nonumber \\
&&
\left.
-\ln\left[ \frac{M_{n}^2}{M_{n}^2-(p+It)^2} \right]
\right\}
\; ,
\end{eqnarray}
in which we impose the condition of $M_{n}^{2}>(p+It)^2$. If $M_{n}^{2}<(p+It)^2$, we obtain
\begin{eqnarray}
\Xi_{I,n}^{(f)}(p)
\!&=&\!
\frac{3\alpha}{8\pi}\left\{\frac{2M_{n}}{\sqrt{(p+It)^2}}\tanh^{-1}\left[ \frac{\sqrt{(p+It)^2}}{M_{n}} \right]
\right.
\nonumber \\
&&
\left.
-\ln\left[ \frac{M_{n}^2}{(p+It)^2-M_{n}^2}\right]-i\pi
\right\}
\; .
\end{eqnarray}
The limit of $p^{\mu}\rightarrow 0$ in (\ref{Xifinite1ap}) yields
\begin{equation}
\Xi_{n}^{(f)}(0)
\!=\!
\frac{3\alpha}{8\pi}\left[\frac{2M_{n}}{\sqrt{t^2}}\tanh^{-1}\left( \frac{\sqrt{t^2}}{M_{n}} \right)
-\ln\left( \frac{M_{n}^2}{M_{n}^2-t^2} \right)
\right]
\; ,
\end{equation}
if $M_{n}^2>t^2$, and
\begin{eqnarray}
\Xi_{n}^{(f)}(0)
\!&=&\!
\frac{3\alpha}{8\pi}\left[\frac{2M_{n}}{\sqrt{t^2}}\tanh^{-1}\left( \frac{M_{n}}{\sqrt{t^2}} \right)
\right.
\nonumber \\
&&
\hspace{-0.5cm}
\left.
-\ln\left( \frac{M_{n}^2}{t^2-M_{n}^2} \right)-i\pi\left(1-\frac{M_{n}}{\sqrt{t^2}}\right)
\right] \; , \;\;
\end{eqnarray}
if $t^2>M_{n}^2$.

%
\vspace{0.5cm}

{\bf Data Availability Statement: No Data associated in the manuscript.}

\end{document}